# Multipath Approach for Reliability in Query Network based Overlaid Multicasting

Ashutosh Singh and Yatindra Nath Singh
*Department of Electrical Engineering,
Indian Institute of Technology,
Kanpur-208016, Uttar Pradesh, India
{singhash, ynsingh}@iitk.ac.in*

**ABSTRACT**

In Application layer multicast (ALM) also called Overlay Multicast, multicast-related functionalities are moved to end-hosts. The key advantages, overlays offers, are flexibility, adaptability and ease of deployment [1]. Application layer multicast builds a peer-to-peer (P2P) overlay multicast tree topology consisting of end-to-end unicast connections between end-hosts. End users self organize themselves into logical overlay networks for efficient data delivery. Major concern in designing ALM protocol is how to build and maintain a topology, to route data efficiently and reliably.  We propose here a scheme in which the topology is built incrementally while maintaining dual feeds of the media stream to any node from the source with minimum differential delay in receiving packets from both alternatives. We have made the assumption of availability of a P2P query search network. This enables building of multicast tree directly as an overlay. There is no need of maintaining an overlaid mesh and running multicast routing protocol to maintain a multicast tree in this mesh. Thus the scheme is much more simplified than in the earlier work on multicast overlay mesh creation and management and then creation and management of multicast trees using this mesh.

## I INTRODUCTION

The internet has seen an unprecedented growth due to the success of one-to-one applications such as reliable file transfer, electronic mail and http based information access over web. The most of the Internet's infrastructure is unicast-only, and does not provide efficient support for real time multicasting applications viz. Internet-TV, Video conferencing, live Lecture Delivery Systems (LLDS) and content delivery networks over Internet. In these applications, copies of a message need to be transported to multiple recipients at different locations. IP multicast [2] at network layer defines an efficient way for multicasting whereby the sources transmit only one copy of the data and intermediate multicast enabled routers make the required number of copies for onward transmission. However IP multicast has not been widely deployed due to the various reasons [3]. IP multicast requires routers to maintain per-group state which leads to scaling constraints when numbers of groups become huge. Although, possibly using aggregation techniques, the number of states can be reduced, IP multicast still needs changes at infrastructure level (deployment/ enabling of multicast routing protocols in the routers) in the Internet. Internet lacks in multicast support as it evolved primarily for  unicast applications since beginning. IP multicast is based on best-effort data delivery, and hence cannot support QoS. This is not desirable in applications where real time synchronous data delivery is needed such as multi-party gaming and multi-party conferencing. The QoS is not possible until it is implemented in all the lower layers of the network. Finally IP multicast provides only a limited support for group management, multicast address allocation and network management.  In IPv4, multicast is an optional service as it has matured much later. As a result, most of the networks have not enabled multicast or provide it as only as a value added service. The multicast has been mostly limited to 'islands' of network domains under single administrative control or in local area networks.

Application layer multicast (ALM), also known as Overlay Multicast is an attractive alternative solution where multicast-related functionalities are moved to end-hosts. We can see this as a mechanism to provide multicast services using whatever available unicast network services. The key advantages, overlays offer, are flexibility, adaptability and ease of deployment. Since the multicast connections here are based on end hosts, there is no need of multicast enabled network routers. ALM achieves multicast via piece-wise unicast connections.

A peer-to-peer (P2P) overlay network is needed for ALM implementation. In this P2P network, the distributed index of resources is maintained. For ALM implementation, the resources will typically be end nodes who can potentially act as media forwarders (equivalent of multicast routers). Each node desirous of receiving the media feed searches for the resource in distributed index in P2P overlay, connects to a node having the resource to get the feed, and then adds itself as a resource provider in the distributed index. Once a node is acting as forwarder for sufficient number of node, it can remove itself from the distributed index as now it may not able to accept requests for the feed from the new nodes. It shall be noted that at any point of time, there are two overlays, one for maintaining distributed resource index, also called query network and second multicast data transport network.

However ALM incurs a performance penalty over IP Multicast. Links near the end users carry redundant copy of data and also the delay to some of the end users is more than what would have been in the case of IP multicast. The major concerns in ALM are how to form an efficient topology for reliable media transport. Switching off a single node has a potential to partition the whole overlay multicast network interrupting media feed distributions to some of the subscribing nodes. This problem is prominent in the multicast topology creation mechanism outlined above. Here the overlay itself is a tree, and hence failure of a node will fragment the multicast topology unless the recovery mechanism is built. While, in the mesh based overlays where the multicast tree is created through separately running multicast routing protocols, the mesh itself is created in such a way that a node or link failure does not partition the network. A streaming application usually has a playback deadline by which data delivery and loss recovery have to be accomplished.

To enhance the reliability, we propose here a scheme in which data forwarding topology is incrementally built. The topology governs the data forwarding in such a way that it maintains double feed to any node from the source. These two feeds come from two different node-disjoint paths from the source and thus resilient to single failures in overlay network. For this, we need a query search network with distributed indexing service that maintains the list of presently active potential forwarders for the two different feeds. Distributed indexing service based query search network helps in locating/searching the source for the desired feeds. Peers form a structured overlay for streaming media forwarding. Each node maintains a table of next hop clients to whom media stream has to be forwarded. It also maintains links with parents from which duplicate feeds are received. The algorithm further allows healing of topology after a failure to keep the two feeds intact to every node via two node-disjoint paths from source.

## II RELATED WORK

### A. Scalable overlaid multicast topology creation methods

Shi and Turner [4] took the case of an overlaid multicast network (OMN) that provides multicast services for real-time audio and video streaming applications through a set of distributed MSNs (Multicast Service Nodes), which communicate with hosts and with each other using standard unicast mechanisms. Authors tried to optimize end-to-end delay and MSN interface bandwidth usage at the routing sites. To solve minimum diameter, degree limited spanning tree problem, Compact Tree (greedy) algorithm is used that builds a spanning tree incrementally. To solve limited diameter, residual-balanced spanning tree problem, Balanced Compact Tree algorithm is used which proved effective in achieving a good balance between residual degree and diameter bound.

Banerjee *et. al.* [5] describes a low overhead hierarchical clustering scheme NICE for ALM for low bandwidth real-time data applications with large receiver sets. This protocol is robust in the sense that failure of any number of group members does not affect the other members in the group. Members are assigned to different layers. Hosts in each layer are partitioned into a set of clusters. All hosts are part of the lowest layer $L_0$. The cluster leaders of all the clusters in layer $L_i$ join layer $L_{i+1}$. There are at most $\log_k N$ layers and the highest layer has only a single member. In each cluster of each layer, the control topology is a clique and the data topology is a star.

### B. ALM loss recovery approaches

ALM loss recovery approaches are broadly classified as proactive and reactive. In a proactive approach, redundant packets are sent along with the data packets which can be used to reconstruct the original data in case some data packets are lost. Such approaches include Forward Error Correction (FEC), Digital Fountain [6], Network Coding and layered coding scheme e.g. Multiple Description Coding (MDC) with multi-path transmission. In a reactive approach, lost packet is retransmitted after the loss has occurred and receiver requests for the lost packets. Probabilistic resilient multicast (PRM) [7] includes both proactive and reactive components.

In probabilistic resilient multicast (PRM) enhanced NICE (Nice is the Internet Cooperative Environment) protocol, Banerjee *et. al.* [7] introduced multicast data recovery scheme with two components. A proactive component called randomized forwarding in which each overlay node chooses a constant number of other overlay nodes uniformly at random and forward data to each one of them with a low probability, and a reactive component called triggered NAKs to handle data losses due to link errors and network congestion, have been proposed.

A good loss recovery approach should meet the following objectives [8]: low residual loss rate, low recovery latency, low recovery overhead and low deployment overhead.

## III OUR APPROACH TOWARD RESILIENCE

The work presented in this paper gives an approach towards resilience of live streaming traffic in an application layer multicast network. Our approach and PRM [7] fall in the same category of resilient approaches called multiple tree approach. But there are two main differences. The primary difference is that our approach builds a data forwarding topology based on P2P query network. The second difference is that we deterministically maintain dual tree at all times for dual feed to each node while in PRM redundant feeding is done on randomly selected links. Thus we expect high reliability in our design. The different aspects of our approach are described below.

### A. Location and search for the feed

The distributed index will maintain list of presently active sources for the two feeds $f_1$ and $f_2$ of the same video stream. If there is more than one live stream, each one of them will have two feeds. For example for stream 1, the two feeds are denoted as $1-f_1$ and $1-f_2$; for stream 2, these will be $2-f_1$ and $2-f_2$. Sources or the nodes which are willing to act as forwarders for a feed publish their advertisement in the distributed index in the p2p query network. Using the keyword describing a feed, a new node can find the potential forwarder for a feed from this distributed query network. For example, all the nodes willing to forward $1-f_1$ can be searched by the keyword $1-f_1$. For maintaining distributed index, the query network can be built using any of the distributed P2P lookup protocol viz. Chord, Tapestry or Pastry. The distributed index should maintain a list of minimum number of active forwarders for each of the two feeds. Any query is provided with the best source for each feed. The node directly makes connection with that forwarder after getting the response. Every node gets a feed each from $f_1$ and $f_2$ forwarders. It also puts up advertisement for being potential forwarders for either $f_1$ or $f_2$. The nodes which are already forwarding to sufficient number of nodes can remove their advertisements to avoid their overloading.

### B. Node Design

The node receiving the dual feed of a stream should use buffering to take care of differential delay in the duplicate data feeds. Two feeds will be giving packets with different delays most of the time. In case of failure, and hence interruption of one feed, the playback of the data feed should remain uninterrupted for user. The stream should be buffered before being given to media player such that longer delay feed's packet also arrive while a copy of it is already present in the queue which arrived via shorter delay path. Figure 1 and Table 1 describe the scenario. Buffering ensures that even when a feed gets interrupted, the user gets the playback of multicast transmission without interruption.

### C. Join Process

Any new node who wants to get the feed, will find the list of peers who are willing to provide the feed; with their parent nodes indicated, either from a distributed P2P query network or from a central server called indexing server. Indexing servers keep record of these peers in a table as shown below in Table 2.

| Packet nos. time | from faster feed | from slower feed | media player is playing | stored in the buffer | | | |
|---|---|---|---|---|---|---|---|
| t = 0 | 70 | 66 | 65 | 66 | 67 | 68 | 69 |
| t = -1 | 69 | 65 | 64 | 65 | 66 | 67 | 68 |
| t = -2 | 68 | 64 | 63 | 64 | 65 | 66 | 67 |
| t = -3 | 67 | 63 | 62 | 63 | 64 | 65 | 66 |
| t = -4 | 66 | 62 | 61 | 62 | 63 | 64 | 65 |
| t = -5 | 65 | 61 | 60 | 61 | 62 | 63 | 64 |

Table 1: An example to show the working of buffer at a node

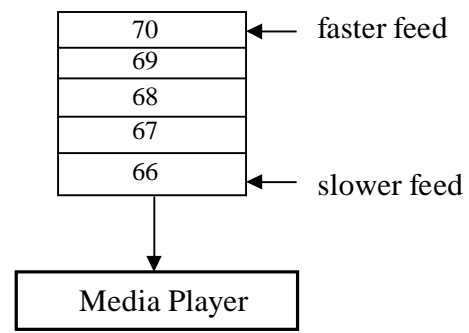

Figure 1: Single buffer created for two feeds from which media player plays the stream

| Peer ID | Degree | If willing to be | If willing to be |
|---|---|---|---|

|  |  | Source for f-1, its parent for f-1 | Source for f-2, its parent for f-2 |
|---|---|---|---|
|  |  |  |  |

Table 2: Record of peers at an indexing server

In order to maintain dual feed which is node and path disjoint, each video is streamed as two identical feeds. These are called as feed-1 (f-1) and feed-2 (f-2). Each new node chooses two peers, one for getting the f-1 and the other one for getting the f-2. Any node, except the source, will get two feeds. It will also advertise itself as a peer willing to relay either f-1 or f-2 only. This will be done by the node by registering itself in indexing server or in distributed P2P Query network.

Usually, the peers getting the two feeds will use them to create a single buffer from which the media player will play the stream. When the number of outgoing relayed feeds reaches maximum, the peer will remove itself from the list of peers willing to relay the feed. This will be done by removing the advertisement from P2P query network or by sending a message to indexing server.

### D. Removal of node from multicast tree

When a member leaves gracefully, it informs all of its children in advance and these children in turn inform further to their children and so on. In this way the sub tree rooted at the leaving node are informed about the departure. Any node in this sub tree, who may have advertised itself as source of feed in distributed query network, will un-publish the advertisement. Now the node who is not getting the feed i.e. next immediate children of leaving node will find new feed source and attaches to them. Once the attachment is done, each such node can inform the same to sub tree rooted at them permitting them to publish the advertisement as potential feed source.

The abrupt departure of any member may cause interruption in playback if sufficient number of packets has not been buffered since it may take time to notify the failure to each node in sub tree and eventually to resume the removed feed from an alternate source.

An easiest way to get the removed feed is to get it from the departing node's parent. We recall that, during join process, when a node gets sources for f-1 and f-2, it also gets information about its grandparent. This grandparent node information can be used now and feed deprived nodes can get the feed from their grandparent. To avoid flash crowd, grandparent node can suggest requesting nodes to connect to its parent or grandparent.

### E. Data Feed Management example

Let us consider an example how such an overlay gets created. Let S be a source of the media stream. When node 1 comes to know about it, it communicates with S directly to get the feed. The S tells it that this feed is f-1. Thus node 1 registers in query network or indexing server as a possible relay source for f-1. When node 2 comes alive, it finds that S is the source for feed, and 1 is possible relay for f-1. It communicates with S to get a direct feed. S informs him that it is feed f-2. Node 2 also communicates with node 1 to get f-1. Node 2 registers in query network as possible relay source for f-2. Node 1 will also periodically check if source for feed f-2 has come up or not from query network or indexing server. As soon as node 1 gets informed that node 2 can relay f-2, it connects to node 2 for f-2.

An example of double feed distribution with 10 nodes (including source) is shown below (refer Figure 2 and Table 3) assuming no failure occurs till first 10 nodes connect to the overlay.

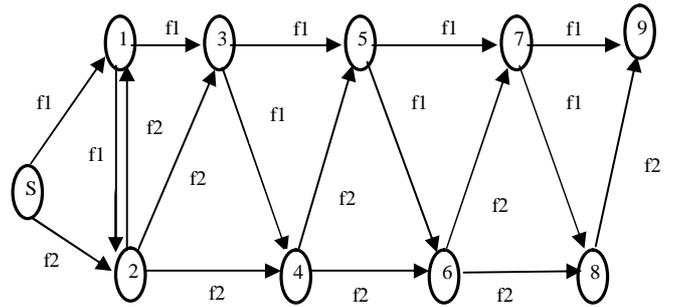

Figure 2: Double feed distribution with 10 nodes

| Node | In-deg | Out-deg | Hop-wise difference in two feeds it gets | What feed it forwards | Its Parent for f1,f2 |
|---|---|---|---|---|---|
| S | --- | 2 | --- | f-1, f-2 | self |
| 1 | 2 | 2 | 1 | f-1 | S, 2 |
| 2 | 2 | 3 | 1 | f-2 | 1, S |
| 3 | 2 | 2 | 0 | f-1 | 1, 2 |
| 4 | 2 | 2 | 1 | f-2 | 3, 2 |
| 5 | 2 | 2 | 0 | f-1 | 3, 4 |
| 6 | 2 | 2 | 1 | f-2 | 5, 4 |
| 7 | 2 | 2 | 0 | f-1 | 5, 6 |
| 8 | 2 | 1 | 1 | f-2 | 7, 6 |
| 9 | 2 | 0 | 0 | f-1 | 7, 8 |

Table 3: Double feed distribution with 10 nodes

The information about a node (viz. its in-degree, out-degree, and its parents for both feeds) are informed to all immediate children when any children joins or anyone of this entity is updated.

### F. Handling event of failure of a node

In case a node in the multicast tree fails, it may interrupt data streaming in f-1 or f-2 tree. This node failure situation can be handled individually for f-1 or f-2 tree by applying one of the following solutions independently of any feed type (f-1 or f-2):

**Solution-1:** In case of failure of a node (N), all its children i.e. immediate children ($N_{g1}$ nodes), grand children ($N_{g2}$ nodes) and all the nodes in the sub tree rooted at failed node will stop getting the feed. Before this failure occurs, each of the node in the sub-tree rooted at failed node (N) were registered as potential source for the feed and after failure of root-node of this sub tree, all these get deprived of the feed. First of all, the $N_{g1}$ nodes detect that their parent has failed and they start searching for a new source. To avoid the possibility that any node in the sub tree, which was registered as potential source; become new parent of $N_{g1}$ nodes, immediately after detecting the failure, root node of the subtree of $N_{g1}$ nodes broadcast a message to every node in the subtree to not to respond to feed request from any nodes for the transition time duration.

**Solution-2:** In case of failure of a node (N), all of its children i.e. immediate children ($N_{g1}$ nodes), grand children ($N_{g2}$ nodes) and all the nodes in the sub tree rooted at failed node explicitly un-publish their advertisement as possible source from the query network. And once $N_{g1}$ nodes search and connect to a new source and data streaming resumes, the $N_{g1}$ nodes and thereafter the nodes in the lower level of the sub tree can again publish advertisement as a possible source.

**Solution-3:** In case of failure of a node (N), all its children which earlier were kept in DHT tables to act as a possible source are kept in the DHT but an INE (ineligible) tag is applied for estimated transition duration (say 30 units of time) to make them ineligible to act as a source for that duration. This will avoid chances of looping as $N_{g1}$ nodes cannot select inactive sources.

### G. Maintaining out-degree after multiple node failures

Initially when the overlay is formed as discussed above, most of the nodes have their in-degree = out-degree = 2. Though this uniform degree distribution gets disturbed after multiple node failures. We need to maintain the out-degree below a certain maximum value.

### H. Delay optimization in data distribution tree

With time, the data distribution tree may loose uniform distribution of load i.e. some portions of the tree may be found denser as compared to others due to which some nodes might be facing inordinate latency, while there may be free out-degree in the upper part of the tree. The following periodic optimization algorithm tries to minimize the average latency in overall tree.

To reduce average latency in overall tree, nodes at lower level should be shifted toward root if there is a free out degree in a upper level node and if there is an advantage in doing so. We define cumulative children of a node i ($CC_i$) as the total number of children nodes in the sub-tree rooted at this node. At any level, all the nodes in that level tell their value of $CC_i$ to their parent. Any node periodically sends its $CC_i$ value to its parent node. Parent node on the basis of $CC_i$ values obtained from all its children, decides about modification in the tree structure.

Firstly, if a node has free out-degree, it fills that out-degree with one of its grandchildren. The node with free out-degree requests to all of its children to send the node ID and $CC_i$ value of their child with highest $CC_i$ value. The grandchild with highest CCi value is selected out of this set and offered to fill that free out-degree at its grandparent.

Secondly, if father node does not have free out-degree, a node can exchange its position with its children. A node i become eligible to replace its parent if it satisfies the following condition:

$$cc_i > \sum_{j=all\ siblings\ of\ i} (cc_j + 1)$$

## IV PERFORMANCE EVALUATION

The performance metrics to measure the efficiency of our approach are the following: Buffer space required for uninterrupted playback, upper bound of differential delay in the two feeds, Time to detect and repair node failure and partition.

Through simulation we can find probability distribution curve of hop difference in two feeds at certain leaf nodes can be found. We can also observe the network behavior as network grows with different growth rates through simulation.

## V CONCLUSION AND FUTURE WORK

The approach toward maintaining dual-feed is suggested. We believe that in our approach there are components for dual feed maintenance and delay optimization in data distribution tree.

Our observation is that the outbound degree has to have some minimum value also. If a node only receives and do not allow transmissions, it is like free riding. Another observation is that we can have another modification. Let each node advertise its distance from root node. When a node gets multiple options to get the new feed, it can choose the parent on hop count basis.